\documentclass[12pt,preprint]{aastex}

\shorttitle{Pre-eclipse Observations $\epsilon$ Aur}
\shortauthors{Stencel et al.}

\begin{document}


\title{Interferometric Studies of the extreme binary, $\epsilon$ Aurigae: \\ 
Pre-eclipse Observations}


\author{R. E. Stencel}
\affil{Dept. Physics and Astronomy, University of Denver, Denver CO 80208}

\author{Michelle Creech-Eakman}                
\affil{Dept. Physics, New Mexico Tech, 801 Leroy Place, Socorro NM 87801}

\author{Alexa Hart}                
\affil{Dept. Physics and Astronomy, University of Denver, Denver CO 80208}

\author{Jeffrey L. Hopkins}                
\affil{Hopkins Phoenix Observatory, Phoenix AZ 85033}

\author{Brian Kloppenborg}
\affil{Dept. Physics and Astronomy, University of Denver, Denver CO 80208}

\and

\author{Dale E. Mais}
\affil{Palomar Community College, Valley Center CA 92082}



\begin{abstract}
We report new and archival K-band interferometric uniform disk diameters obtained with
the Palomar Testbed Interferometer for the eclipsing binary star
$\epsilon$ Aurigae, in advance of the start of its eclipse in 2009.  The
observations were intended to test whether low amplitude variations in
the system are connected with the F supergiant star (primary), or with the
intersystem material connecting the star with the enormous dark disk (secondary) 
inferred to cause the eclipses.  Cepheid-like radial pulsations of the F star are not detected, nor do we find evidence for proposed 6\% per decade shrinkage of the F star.  The measured 2.27 +/- 0.11 milli-arcsecond K band diameter is consistent with a 300 solar radius F supergiant star at the Hipparcos distance of 625 pc.  These results provide an improved context for observations during the 2009-2011 eclipse.
\end{abstract}


\keywords{techniques: interferometric, stars: atmospheres, binaries:
eclipsing, stars: fundamental parameters}



\section{Introduction}

The prevailing hypothesis concerning the nature of the long period
eclipsing binary FK5 183 (HD 31964, $\epsilon$ Aurigae) features an F
type supergiant star and a putative B star binary - deeply embedded in a
dark, massive, 20 AU diameter cold disk (475K; Carroll, et al. 1991). In
the high mass model, total system mass is inferred to be approximately
29 solar masses, with an orbital separation of 27.6 AU and eclipse
period of 27.1 years (cf. Stencel, 1985).

Flat-bottomed eclipses of two years duration and 0.75 mag depth
optically, suggest that the cold disk covers half the surface area of
the F star (Huang, 1965).  The next eclipse is predicted to start in
2009 August.  Kemp et al. (1986) analyzed polarimetry of the 1984
eclipse and argued that the disk is inclined 2 to 5 degrees from its
orbital plane.  Taken together with a central eclipse brightening that
has varied over the past 3 eclipse events, disk tilt could signal
precession of the disk orientation.  However, the F star outshines the
cold disk by an enormous factor, adding to the mystery of the secondary
itself.  Low amplitude, 67 day quasi-periodic light variations mask the
relative contributions of F star and disk in the pre-eclipse light curve
(Hopkins and Stencel, 2007), and these light variations appear to have
sped up from 89 days over the past few decades (Hopkins, Schanne and Stencel, 2008).
Concurrently, the length of eclipse phases has been changing, eclipse to eclipse.

\subsection{Goals}

The key question to be addressed with new observations is whether
the quasi-periodic 0.1 magnitude variations in V-band light outside of
eclipse are due to F supergiant pulsation - or - due to components
associated with the disk and mass transfer (Stencel 2007).  

The V band $\sim 0.1$ magnitude quasi-periodic variations indicate $\sim 10\%$ luminosity changes in the system.  If these originate in F star changes in temperature or radius, they would amount to of order 5\% in radius, and half that amount or less in temperature terms.  Asteroseismic observations such as those possible with MOST or COROT, along with high dispersion spectroscopic monitoring of line profile variations, should be pursued to explore which parameters are in play.  Interferometry provides a potentially more direct test of diameter variations, given interferometric diameter variation measurement successes with Cepehids like $\zeta$ Gem with PTI (Lane et al. 2000, 2002) and  
$\delta$ Cep and $\eta$ Aql with NPOI (Armstrong et al. 2001), wherein radial variations of up to 6\% (a range of 0.20 +/- 0.03 milli-arcsecond, hereafter, mas) were reported.  If physical variations of the F star in the $\epsilon$ Aur system can be demonstrated to be the cause of, or excluded from causing out-of-eclipse light variations, study of the disk-shaped companion can be more precisely pursued.  This includes interferometric imaging that can determine whether the dark disk in the Huang model actually will be seen against the F star disk.

Adopting the Hipparcos parallax distance of 625 pc for $\epsilon$ Aur, the maximum apparent orbital separation is 44 milli-arcsec (mas), and the F supergiant itself, if 200 solar radii, should subtend $\sim 1.5$ mas.  The reported NPOI diameter of 2.18 mas \citep{nor01} for $\epsilon$ Aur implies a diameter of 290 solar radii at 625 pc.  This is significantly larger than the Cepheid diameters mentioned above and the VLTI/AMBER diameter, 142 solar radii, for the F0 supergiant Canopus, reported by Dominicano de Souza, et al. (2008).  In any event, a 5\% or larger radial change in $\epsilon$ Aur amounts to at least 0.14 mas, which is well within the 0.03 mas error limit possible with current 100 m baseline interferometers.  In addition, the baseline data provided by such observations provides an important dataset against which future in-eclipse observations will be compared.  Thus, we provide this Letter reporting on the status of interferometric data related to the $\epsilon$ Aurigae system.

\section{Observations}


We proposed to use the Palomar Testbed Interferometer (PTI, Colavita et al. 1999)  
in Visibility amplitude mode, K-band, to monitor $\epsilon$
Aurigae during the winter 2007/08 season, on a once per month basis. The
initial observing was conducted on 2007 October 18-20. Calibrators used
and cross-calibrator checks are shown in Tables \ref{tblCalibrators} and \ref{tblCrossCalibration}, selected and vetted following processes described in van Belle et al. (2008).  PTI's K-band
K-low capability over 5 wavelength channels presented an exceptional opportunity to
precisely measure the angular diameter of the primary star in $\epsilon$
Aur.  In order to obtain accurate visibility readings from the
calibration software, one must accurately select calibrators.  In
addition to having well-known coordinates, proper motion and parallax, 
calibrators must be bright enough to be tracked by PTI, appear
point-like in nature (for PTI, $\theta \lesssim 0.8$ mas is suitable
(van Belle et al. 2007, 2008), and have nearly constant visibility
measurements.  Seeing and instrumental issues provide omnipresent limitations that influence the estimated errors on diameter measurements (see below).

In addition to new observations, the PTI archives included several prior measurements which help establish a longer term baseline and check on trends.  Ancillary data on $\epsilon$ Aur includes optical photometry, H $\alpha$ and Spitzer IRS spectra and MIPS data, as part of an observational monitoring campaign (Hopkins, Schanne and Stencel 2008; see also Stencel 2007). 

\section{Data Reduction and Analysis}

PTI data products consist of several levels of data.  Raw data from the
interferometer are called Level 0 data files.  At the end of the
observing night, a program called $vis$ - see Colavita (1999b) - processes
the Level 0 data and creates Level 1 data files.  This data is provided
to the end user as a series of ASCII or FITS files for further
processing. 

Level 1 data consists of Wide-band visibility squared ($V^2$) data,
Spectrometer $V^2$ data, a baseline model, reduction configuration
information, an observer log, a nightly report, the catalog (schedule)
file, and postscript plots of the wide-band and spectral data.  This
information, along with a calibration script and a baseline model
(.baseline file) is processed using two programs contained in the
$V2calib$ package to create calibrated wide- and narrow-band $V^2$ data.

The $V2calib$ package contains the source code for the wide- and
narrow-band calibration programs, $wbCalib$ and $nbCalib$.  After being
compiled, these two programs automate a majority of the data reduction
process by computing calibrated $V^2$ measurements as well as other
ancillary data including $u$- and $v$-projections (spatial frequencies)
for each calibrated scan.  If one does not have a Linux-based system on
which the programs may be compiled, one may also use the Michelson
Science Center's web-based calibration tool, webCalib to produce the
same data products.

Even though the $V2calib$ programs do much to simplify the data
analysis, one cannot be guaranteed to obtain $V^2$ data that is
reasonable without further analysis.  Examining the calibrator-derived system
visibilities helps verify that this exceeds an ideal average better than 0.5, and varies
smoothly over the observing night (see details at the Michelson Science
Center website).  Only two nights, 2007 Oct 19 and 1998 Nov 25 are ideal in terms of the highest system-visibility requirements.  As can be seen in Table \ref{tblAngularDiameters}, the derived angular diameters for these two dates agree within the errors, 2.19 +/- 0.06 mas and 2.25 +/- 0.08 mas, respectively.  Lane et al. (2002) provide a clear discussion of errors in PTI data reduction, and our errors scale with the number of scans reported in their Tables 3 and 4.

We also elected to consider new and archival data points with lower calibrated system
visibilities (down to $\sim$0.2), as long as the nightly system visibility varied
smoothly with time.  After initial results using default settings, we also switched off the ratio correcting feature of the software to achieve more uniform 
results, as recommended by Rachel Akeson at MSC.  
In addition to system visibility requirements, one also needs to
evaluate the performance of the system over an observing night.  One
measure of system performance can be found by cross-calibrating the calibrators.  
Doing this is as simple as running the $V2calib$ programs with a calibration star specified as a target.  Of course, this requires that the data set contains multiple
calibrators during an observing night, and that there are a sufficient
number of data points for the $V2calib$ programs to process into
meaningful data.  Because all of our calibrators are selected to be unresolved (angular
diameters $\theta < 1.0$ [mas]), we expect to obtain $V^2$ values close
to unity.  The results of cross-calibration are summarized in Table \ref{tblCrossCalibration}, where we see that several recent nights approach this criterion.  
Unfortunately, most of the nights with archival data did not contain more than one calibration star.  

After the data reduction, the $V^2$ data and its errors are then fit to a model.  We
elected to use the Uniform Disk (UD) model in which: 

\begin{equation}
V^2 = \frac{(2 J_1(\pi \theta B / \lambda ))^2}
{{(\pi \theta B / \lambda)}^2}
\end{equation}

\noindent where $J_1$ is the first-order Bessel function (approximated
using the first-seven terms of the power-series expansion), $B$ is the
projected baseline ($\sqrt{u^2 + v^2}$), $\theta$ is the stellar angular
diameter in radians, and $\lambda$ is the wavelength of light at which
the data was obtained.  Given the limited data set, we did not pursue more elaborate models for the source size, at this time.

Because this function is non-linear, we elected to create a lookup
table.  This table consisted of values of $(\pi \theta B / \lambda)$
from $0.9$ to $2.36$ (inclusive) in $0.00002$ step increments and their
corresponding $V^2$ values.  Using this method, we were able to match
the $V^2$ readings from PTI with the $V^2$ values in our table to within
$2 \times 10^{-5}$.  After a $V^2$ match was obtained,
we used the corresponding $(\pi \theta B / \lambda)$ value to solve for
the angular diameter.  Using this method, we calculated the theoretical
error in angular diameter that would result from a $0.00002$ increment
in $(\pi \theta B / \lambda)$ to be $8 \times 10^{-16}$ at a
maximum.  Take note that this is several orders of magnitude below any
error that arises from $\Delta V^2$ measurements, e.g. seeing.  The errors on measurements reported here are seeing dominated and future observations need to take care to include a larger number of scans and cross-calibrator measurements to reduce overall uncertainties.





\section{Discussion}

The error-weighted mean K-band uniform-disk angular diameter for $\epsilon$ Aurigae derived from 12
nights between 1997 and 2008 at the Palomar Testbed Interferometer is 
2.27 +/- 0.11 mas.  These values are consistent with the published NPOI and earlier Mrk III 
optical band values of (UDD) 2.18 +/- 0.08 mas and (LDD) 2.17 +/- 0.03 mas (Nordgren et al. 2001), although arguably slightly larger at K-band compared to these optical-band results.  Knowledge of the optical light curve was provided by UBV photometry obtained in parallel at Hopkins Phoenix Observatory - see Hopkins et al. (2008).   No clear correlation could be seen among the limited variations
in the derived diameters and the optical light curve, to the limits imposed by
the measurement errors.  The majority of diameters spanning the longest timespan were measured on a
(nearly) N-S baseline.  We note that Kemp et al. (1986) indicated a polar
axis position angle for the F star of 5 to 45 degrees, and our few N-W baseline points may appear slightly larger on that axis.  We also checked for luminosity-related changes.  
The V magnitude during 2007 Oct
(RJD 4393-6) was 3.035, but by 2008 Feb (RJD 4515) had reached V = 2.98,
an unusually bright maximum, even though the 67 day phasing suggested that a
minimum should have occurred then.  That latter epoch also featured an
usually asymmetric H-alpha profile, with a strong blue emission wing and
redshifted absorption core.  However, the diameters appeared similar
(albeit on a N-W baseline then) and a Mimir spectrum obtained shortly
afterward did not show any significant changes to the weak Brackett
emission, however (see Clemens et al. 2008 - Fig.14).  Additional baseline coverage might reveal azimuthal changes, perhaps associated with proposed equatorial rings \citep{kem86}. 

After the 1984 eclipse ended, Saito and Kitamura (1986) provided
evidence that the F supergiant star was shrinking at a rate of 16\% 
eclipse to eclipse (27.1 years), based on changing duration of eclipse totality during the past few eclipses, assuming the disk was invariant.  At face value, this would result in a decrease of angular
diameter of the F star by nearly 6\% over the 10 year PTI interval reported here.  Within the
dispersion of PTI measurements, we do not confirm any decrease of this magnitude, or have
evidence for significant changes in diameter over the past 10 years, assuming the older PTI, NPOI and/or Mrk III data do not have systematics relative to the more recent measurements.  The eclipse to eclipse variations may be due instead to secular changes in the dark companion object rather than the F star - a point testable with the next eclipse.  The 2.27 mas angular size reported here, when combined with the 625pc Hipparcos distance, implies a primary star diameter of 308 solar diameters.  This is larger than the classically derived diameter for an F0 Ia star (200 solar diameters, Schmidt-Kaler 1965; Allen ApQ 4th Ed.), suggesting the star is possibly cooler than F0 and/or has an extended atmosphere due to the binary interaction.  What is needed are new classification spectra of $\epsilon$ Aurigae, as well as a careful determination of effective temperature from a spectral
energy distribution study. 

Further progress in the study of $\epsilon$ Aurigae should be possible
by applying interferometric imaging to the eclipse event during
2009-2011.  If the Huang model is basically correct, the passage of a
dark disk, bisecting the F star surface, should produce a
straightforward change in the fringe patterns - from circular symmetry
of a single disk, to an asymmetry from a close pseudo-binary star pair
of bright limbs during totality, modulo pulsation phenomena.  We ask
observers with suitable resources to make this star a priority for
frequent observation during this rare opportunity.



\acknowledgments

We are pleased to thank the staff at Palomar Observatory for their
hospitality, for encouragement from Michelson Science Center key people
Rachel Akeson, Kevin Rykowski, Gerard van Belle and the PTI Collaboration.  We thank Dan Clemens and April Pinnick for obtaining Mimir spectra of $\epsilon$ Aur, and Ed Guinan for insightful comments, and the referee for careful reading and helpful suggestions on the manuscript.  We are grateful for support of this effort in part from JPL/Spitzer award 1275955 and from the bequest of 
William Herschel Womble in support of astronomy at the University of Denver.



{\it Facilities:} \facility{Palomar Observatory: Palomar Testbed Interferometer}

\clearpage
\begin{table}[ht]
\tabletypesize{\scriptsize}
\caption{Calibrators for HDC31964 used during observations. \label{tblCalibrators}}
\begin{tabular}{llllllll}
\hline
\hline
Star Name & RA          & DEC         & $\mu_{RA}$ & $\mu_{Dec}$    & Parallax   & UDD  & Error \\
	  &             &             &            &                & (Hipparcos) & [mas] & [mas] \\	
\hline 
HD 23838 & 3 50 04.420 	& +44 58 04.28	& -0.03780 & -0.02682	& 0.00941	& 0.877	& 0.055 \\
HD 29203 & 4 38 05.877 	& +46 14 01.15	& 0.02569  & -0.02157 	& 0.00568 	& 0.587 & 0.102	\\
HD 29645* & 4 41 50.256 	& +38 16 48.65	& 0.24153  & 0.09788	& 0.03203	& 0.542	& 0.009 \\
HD 30138 & 4 46 44.478	& +40 18 45.33	& 0.00899  & -0.0371	& 0.00736	& 0.784	& 0.047 \\
HD 30823* & 4 52 47.757	& +42 35 11.85	& -0.01107 &  0.00011	& 0.00631	& 0.280	& 0.027 \\
HD 32630* & 5 06 30.892 	& +41 14 04.10	& 0.03060  & -0.06841 	& 0.01487 	& 0.374 & 0.079 \\
HD 34904 & 5 22 50.314 	& +41 01 45.33	& -0.01249 & 0.00294	& 0.01087	& 0.339	& 0.021 \\
\hline
\end{tabular}
\end{table}
*from van Belle et al. (2008).
\clearpage

\begin{table}[ht]
\caption{Diameters obtained from Wide-Band Visibility mode data. \label{tblAngularDiameters} }
\begin{tabular}{lllllllll}
\hline
\hline
UTDate, & GMT start & Baseline* & Nscan & Mode & V$^2$ & UDD   & Error & V \\
JD-2,450,000 &     &          &  sets** &      &        & [mas] & [mas] & [mag] \\
\hline
2007Oct19, 4393	& 09:57	& NS & 14 & K-low & 0.516 & 2.19 & 0.06 & 3.036 \\ 
2007Oct20, 4394  & 10:21      & NS & 6 & K-high & 0.544 & 2.16 & 0.12 & 3.036 \\ 
2007Oct21, 4395	& 10:45	& NS & 3 & K-low 	& 0.583 & 1.90 & 0.13 & 3.036 \\ 
2007Dec23, 4458	& 04:41	& NW & 6 & K-low 	& 0.574 & 2.36 & 0.14 & 3.046 \\ 
2007Dec24, 4459	& 04:48	& NW & 6 & K-low 	& 0.565 & 2.37 & 0.11 & 3.043 \\ 
2008Feb16, 4513	& 03:05	& NW & 2 & K-low 	& 0.527 & 2.60 & 0.15 & 2.98 \\ 
2008Feb17, 4514	& 04:48	& NW & 5 & K-low 	& 0.572 & 2.28 & 0.15 & 2.98 \\ 
2008Feb18, 4515	& 03:01	& NW & 5 & K-low 	& 0.624 & 2.25 & 0.12 & 2.98 \\ 
\hline
Archival Data & & & & & & & & \\
\hline
1997Oct22, 0744	& 11:54	& NS 	& 1  & K-low & 0.376 & 2.50 & 0.17 & 2.986 \\ 
1997Nov09, 0762	& 09:38	& NS	& 2	& K-low & 0.438 & 2.32 & 0.09 & 2.977 \\ 
1998Nov07, 1125	& 10:25	& NS	& 4	& K-low & 0.515 & 2.09 & 0.10 & 2.997 \\ 
1998Nov25, 1143	& 10:19	& NS	& 2	& K-low & 0.458 & 2.25 & 0.08 & 2.998 \\ 
\hline
1998Nov26, 1144	& 10:20 	& NS	& 1	& \multicolumn{3}{l}{No Cal Stars} & & 2.998 \\
2005Dec11, 3715	& 06:33	& NW	& 1	& \multicolumn{3}{l}{No Cal Stars} & & 3.02 \\
2006Jan31, 3766	& 04:27	& NW	& 83	& \multicolumn{3}{l}{No Cal Stars} & & 3.08 \\
\hline
\end{tabular}
\end{table}
*N-S baseline, 109 meters; N-W baseline, 86 meters.\\
**Each Level 1 scan set consists of 2 or more integrations of 25 sec each during which fringe visibility was averaged (http://msc.caltech.edu/software/PTISupport/v2/sum.html ).

\clearpage
\begin{table}[ht]
\caption{Cross-Calibrator Visibility squared measurements. \label{tblCrossCalibration}}
\begin{tabular}{llllll}
\hline
\hline 
Date & Star Name & $\langle$Cal $V^2 \rangle$ & $\langle$Error$\rangle$ & $\langle$ Sys $V^2 \rangle$ & $\langle$Error$\rangle$\\
\hline 
2007Oct19	& HD29645	& 0.95 	& 0.05 & 0.66 & 0.02 \\
2007Oct19	& HD29203	& 0.95	& 0.05 & 0.66 & 0.01 \\
2007Oct20	& HD30138	& 0.87	& 0.06 & 0.46 & 0.02 \\
2007Dec23	& HD30138	& 0.90	& 0.08 & 0.30 & 0.02 \\
2008Feb18	& HD30138	& 0.67	& 0.08 & 0.37 & 0.02 \\
\hline
\end{tabular}
\end{table}

\end{document}